\documentclass[11pt]{article}
\usepackage{amsmath, amssymb, amsthm,bbm}
\usepackage{algorithm, algorithmic}
\usepackage{geometry}
\RequirePackage[colorlinks,citecolor=blue]{hyperref}

\usepackage{setspace}

\usepackage{algorithm}
\usepackage{algorithmic}

\RequirePackage[numbers]{natbib}

\usepackage{orcidlink}

\theoremstyle{plain}
\newtheorem{theorem}{Theorem}
\newtheorem{assume}{Assumption}
\newtheorem{lemma}{Lemma}

\usepackage{authblk} 

\title{\bf Censored Graphical Horseshoe: Bayesian sparse precision matrix estimation with censored and missing data 
}

\author[1]{The Tien Mai\orcidlink{0000-0002-3514-9636}}
\author[2]{Sayantan Banerjee \orcidlink{0000-0001-5414-4817}}

\affil[1]{\small Norwegian Institute of Public Health, Oslo, 0456, Norway.
}

\affil[2]{\small OM \& QT Area, Indian Institute of Management Indore, M.P. 453556, India.
}

\date{\small Email: \texttt{the.tien.mai@fhi.no}, 
\\
\texttt{sayantanb@iimidr.ac.in} 
}

\begin{document}
	
\maketitle

	\begin{abstract}
Gaussian graphical models  provide a powerful framework for studying conditional dependencies in multivariate data, with widespread applications spanning biomedical, environmental sciences, and other data-rich scientific domains. While the Graphical Horseshoe (GHS) method has emerged as a state-of-the-art Bayesian method for sparse precision matrix estimation, existing approaches assume fully observed data and thus fail in the presence of censoring or missingness, which are pervasive in real-world studies. In this paper, we develop the Censored Graphical Horseshoe (CGHS), a novel Bayesian framework that extends the GHS to censored and arbitrarily missing Gaussian data. By introducing a latent-variable representation, CGHS accommodates incomplete observations while retaining the adaptive global–local shrinkage properties of the Horseshoe prior. We derive efficient Gibbs samplers for posterior computation and establish new theoretical results on posterior behavior under censoring and missingness, filling a gap not addressed by frequentist Lasso-based methods. Through extensive simulations, we demonstrate that CGHS consistently improves estimation accuracy compared to penalized likelihood approaches. Our methods are implemented in the package \texttt{GHScenmis} available on Github: \url{https://github.com/tienmt/ghscenmis}.
	\end{abstract}


    \sloppy
	\section{Introduction}
	
	Gaussian graphical models (GGMs) are a central tool for modeling conditional dependencies among multivariate random variables, with applications in fields ranging from genomics and neuroimaging to finance \citep{pourahmadi2011covariance,ryali2012estimation,fan2016overview,callot2021nodewise}. In GGMs, the precision matrix $\Omega = \Sigma^{-1} \in \mathbb{R}^{p \times p} $ encodes the conditional independence structure: a zero in an off-diagonal entry indicates conditional independence between the corresponding nodes \citep{lauritzen1996}. Estimating sparse precision matrices is particularly important in high-dimensional settings, where the number of variables $p$ may be comparable to or larger than the sample size $n$, and sparsity can be leveraged to recover the underlying graph structure. Popular frequentist approaches include the graphical lasso \citep{friedman2008sparse}, graphical SCAD \citep{fan2009network}, and CLIME \citep{cai2011constrained}, which provide consistent estimators of high-dimensional inverse covariance matrices under various sparsity patterns \citep{cai2016estimating,jankova2017honest}.
	
	Bayesian methods for precision matrix estimation offer adaptive regularization and uncertainty quantification, but work in this area is relatively limited.  \cite{wang2012bayesian} proposed a Bayesian graphical lasso via block Gibbs sampling, while  \cite{banerjee2015bayesian} introduced a prior combining a continuous shrinkage component with a point-mass at zero for off-diagonal entries, deriving posterior contraction rates and Laplace approximations for model selection. Spike-and-slab variants with double-exponential priors were explored by  \cite{gan2019bayesian}, 
	and  \cite{atchade2019quasi} used a quasi-likelihood approach with favorable posterior contraction. A limitation of spike-and-slab priors is the presence of binary indicators, which can impede posterior exploration.
	
	The Graphical Horseshoe (GHS) prior \citep{li2019graphical} overcomes this challenge by applying global-local shrinkage to precision matrices, providing strong adaptive regularization and superior empirical performance compared to both frequentist and Bayesian alternatives. Recent work has further explored variants of the Horseshoe prior, including pseudo-likelihood approaches \citep{zhang2022contraction} and Horseshoe-like priors \citep{sagar2024precision}.
	
In practice, data often include censored or missing values, particularly in biomedical, environmental, or epidemiological studies where measurements may fall below detection limits, 
see for example \cite{derveaux2010successful,mccall2014non,psaila2016single,pipelers2017unified}. 
Ignoring censoring or missingness can lead to biased inference, yet methods to handle such data within high-dimensional Bayesian GGMs remain scarce. Early work on covariance estimation with missing or censored data was developed by  \cite{lounici2014high,pesonen2015covariance}, but these methods were not Bayesian. Frequentist approaches for censored graphical lasso have been proposed in \cite{augugliaro2020penalized, stadler2012missing}, with software implementations available in \cite{augugliaro2023cglasso}, but Bayesian alternatives remain underdeveloped.
	
In this paper, we extend the Graphical Horseshoe to handle censored 
and arbitrarily missing Gaussian data. 
Our censored Graphical Horseshoe (CGHS) model introduces a latent-variable formulation to account for censored observations and missingness, while leveraging the adaptive shrinkage properties of the Horseshoe prior to recover sparse precision matrices. This approach allows robust inference of conditional dependencies in challenging real-world datasets where standard GHS methods may fail.

To perform posterior inference for our model, we have developed a Gibbs sampler that updates a block of parameters and latent variables in each step. The sampler cycles through several key steps: first, it imputes censored latent values or imputes the missing data; second, it updates the regression parameters and residual variances; third, it adapts the local and global shrinkage scales; and finally, it reconstructs the precision matrix. This process generates samples from the joint posterior distribution of the model parameters and latent variables, which can then be used to form reliable estimates of the underlying graphical structure. 

We also develop a comprehensive set of theoretical results that characterize the behavior of the posterior distribution in each of the settings we consider. This aspect of our work is genuinely novel, and to the best of our knowledge, no theoretical result has been established for the frequentist methods in complex scenarios involving censoring or missingness. More specifically, we derive posterior concentration rates of the precision matrix for both censored-data and missing-data regimes. These results provide rigorous justification for the stability and reliability of our Bayesian formulation, demonstrating that the method adapts to the underlying signal even under challenging data-generating mechanisms.

In addition to our theoretical contributions, we present an extensive suite of simulation studies designed to assess the finite-sample performance of the proposed approach. Across a broad range of sparsity levels, noise structures, and censoring or missing-data patterns, our Bayesian estimator frequently outperforms its Lasso-based counterparts. Finally, we illustrate the practical utility of our method through an application to single-cell data, where the model successfully captures key biological signals while accommodating the intrinsic sparsity and uncertainty in these datasets. 

The rest of the paper is presented as follows. Section~\ref{sc_censored_GHS} present models and our approach with Gibbs sampler for the case of censored data. Section~\ref{sc_missing} present models and our approach with Gibbs sampler for the case of missing data. Numerical studies with simulations and a real data application are given in Section~\ref{sc_numerical} and Section~\ref{sc_real_Data}. Theoretical results are presented in Section~\ref{sc_theory} with proofs in Appendix~\ref{sc_proofs}. We discuss and conclude in Section~\ref{sc_conclusion}. Additional simulation results and plots are in Appendix~\ref{app:addl_simu_results}.
	
\section{Censored graphical Horseshoe}
	\label{sc_censored_GHS}
	\subsection{Model specification}

 Let $Y = (Y_1, \dots, Y_n) \in \mathbb{R}^{n \times p}$ denote the data matrix, where each $Y_i \sim \mathcal{N}_p(0, \Omega^{-1})$. Here $\Omega$ denotes the precision (inverse-covariance) matrix. In practice, however, the variables are not always fully observed; instead, the measurements are subject to left censoring at variable-specific thresholds. For each coordinate $j$, if the true value $Y_{ij}$ falls below a censoring threshold $c_j$, we only record the threshold itself. Thus, the observed values take the form
\begin{equation}
    \widetilde{Y}_{ij} = 
\begin{cases}
	Y_{ij}, & \text{if } Y_{ij} > c_j, \\
	c_j, & \text{if } Y_{ij} \leq c_j,
\end{cases}
\label{eq_observed_model_sencored}
\end{equation}
for $ i \in \{1,\ldots, n \} $ and  $ j \in \{1,\ldots, p \} $.

To deal with censoring, we introduce latent uncensored variables $Z = (Z_1, \dots, Z_n)^\top \in \mathbb{R}^{n \times p} $ with the same distribution as the original Gaussian model, that is, $Z_i \sim \mathcal{N}_p(0, \Omega^{-1})$. The observed data are then related to the latent process through $\widetilde{Y}_{ij} = \max(Z_{ij}, c_j)$. Inference therefore focuses on the sparse precision matrix $\Omega$ given the censored observations $\widetilde{Y}$.

For each observation $i$, we define the sets 
$$
\mathcal{O}_i = \{j : \widetilde{Y}_{ij} > c_j\}
,
\;
 \text{ and } \;
 \mathcal{C}_i = \{j : \widetilde{Y}_{ij} = c_j\}
$$ to denote the indices of uncensored and censored entries, respectively. Partitioning $Z_i$ accordingly, the observed data likelihood can be expressed as a product of Gaussian density terms for the uncensored components and multivariate Gaussian distribution function terms for the censored ones:
\begin{equation}
\label{eq_likelihood_censored_data}
    L(\Omega; \widetilde{Y}) 
    = \prod_{i=1}^n 
\phi_{|\mathcal{O}_i|}\!\left(\widetilde{Y}_{i,\mathcal{O}_i}; 0, \Sigma_{\mathcal{O}_i, \mathcal{O}_i}\right)
\cdot 
\Phi_{|\mathcal{C}_i|}\!\left(c_{\mathcal{C}_i} \mid \mu_{i,\mathcal{C}_i|\mathcal{O}_i}, \Sigma_{\mathcal{C}_i|\mathcal{O}_i} \right)
,
\end{equation}
where $\Sigma = \Omega^{-1}$, and the conditional mean and covariance of the censored components given the observed ones are
\[
\mu_{i,\mathcal{C}_i|\mathcal{O}_i} = \Sigma_{\mathcal{C}_i, \mathcal{O}_i} \Sigma_{\mathcal{O}_i, \mathcal{O}_i}^{-1} \widetilde{Y}_{i,\mathcal{O}_i},
\]
\[
\Sigma_{\mathcal{C}_i|\mathcal{O}_i} = \Sigma_{\mathcal{C}_i, \mathcal{C}_i} - \Sigma_{\mathcal{C}_i, \mathcal{O}_i} \Sigma_{\mathcal{O}_i, \mathcal{O}_i}^{-1} \Sigma_{\mathcal{O}_i, \mathcal{C}_i}.
\]

\subsection{Prior specification}

To induce sparsity in the precision matrix, as in \cite{li2019graphical}, we place the graphical Horseshoe prior on the off-diagonal entries. For indices $j < k$, the prior specification takes the hierarchical form
\begin{equation}
\label{eq_horseshoe_prior}
\begin{aligned}
     \omega_{jk} \mid \lambda_{jk}, \tau 
  &   \sim 
  \mathcal{N}(0, \lambda_{jk}^2 \tau^2), 
     \\
\lambda_{jk} 
& \sim 
\mathcal{C}^+(0,1), 
\\
\tau 
& \sim \mathcal{C}^+(0,1)
,   
\end{aligned}
\end{equation}
where $\mathcal{C}^+$ denotes the half-Cauchy distribution. The diagonal elements $\omega_{jj}$ are constrained to ensure positive definiteness and are typically assigned noninformative priors or estimated under such constraints.

Each coefficient is assigned its own scale factor, denoted by $\lambda_{jk}$, which acts as a local shrinkage parameter, while a common scale factor, $\tau$, serves as the global shrinkage parameter across all coefficients. This hierarchical prior structure produces two desirable effects: a sharp concentration of prior mass around zero, which aggressively suppresses noise variables, and heavy tails, which allow large, informative signals to remain largely unaffected \citep{carvalho2010horseshoe}.

\subsection{Data augmentation for censoring}

The censored structure naturally lends itself to a data augmentation approach in which the latent Gaussian variables $Z$ are treated as missing data. 
Given observations data $\widetilde{Y}$ and the censoring thresholds $ (c_1, \ldots, c_p) $, the augmented variables $Z_{ij}$ are updated from their conditional distributions. If an observation is uncensored, then $Z_{ij}$ coincides with $\widetilde{Y}_{ij}$ and follows its conditional Gaussian law. If an observation is censored, then $Z_{ij}$ is drawn from the same Gaussian distribution but truncated to the region $(-\infty, c_j]$. Formally, the conditional distribution takes the form
\[
Z_{ij} \mid \widetilde{Y}_{ij}, \Omega, Z_{i,-j} \sim
\begin{cases}
	\mathcal{N}(\mu_{ij}, \sigma_j^2), & \text{if } \widetilde{Y}_{ij} = Z_{ij}, \\
	\mathcal{N}(\mu_{ij}, \sigma_j^2) \, \mathbb{I}(Z_{ij} \leq c_j), & \text{if } \widetilde{Y}_{ij} = c_j,
\end{cases}
\]
where $\mu_{ij}$ and $\sigma_j^2$ are derived from the conditional distribution of $Z_{ij}$ given the remaining coordinates under the multivariate normal model:
\[
Z_{ij} \mid Z_{i,-j} 
\sim 
\mathcal{N}\!\left(-\frac{1}{\omega_{jj}} \sum_{k \neq j} \omega_{jk} Z_{ik}
\,
, \, \omega_{jj}^{-1}\right).
\]

\subsection{Gibbs sampler for censored Graphical Horseshoe}
\label{sc_nodewide_regression}

		
		\subsubsection{Parameterization via nodewise regression }
		
		To parameterize $\Omega$, we adopt the nodewise regression framework \citep{meinshausen2006variable,van2014asymptotically}. Denoting $Z_{\cdot j}$ as the $j$th column of $Z$, and $Z_{\cdot,-j}$ the $n \times (p-1)$ submatrix of $Z$ excluding column $j$, the conditional distribution of $Z_{\cdot j}$ given the remaining variables is modeled as
		\begin{equation*}
			Z_{\cdot j} \,\big|\, Z_{\cdot,-j},\ \theta_j,\ \sigma_j^2
			\;\sim\; \mathcal{N}\big(Z_{\cdot,-j}\,\theta_j,\ \sigma_j^2 I_n\big),\; j = 1,\ldots,p,
		\end{equation*}
		where $\theta_j \in \mathbb{R}^{p-1}$ collects the regression coefficients and $\sigma_j^2 > 0$ is the conditional variance. The connection to the precision matrix is explicit:
		\begin{equation*}
			\Omega_{jj} = \sigma_j^{-2}, \qquad \Omega_{-j,j} = -\sigma_j^{-2}\,\theta_j.
		\end{equation*}
		Since each of the $p$ regressions yields a separate estimate, a symmetrization step (described below) is necessary to reconcile them into a single symmetric precision matrix. 
        In this form, we have that, for  each coefficient $\theta_{jk}$ with $k \neq j$,
		\begin{equation*}
\theta_{jk}\,\big|\,\lambda_{jk}^2,\tau_j^2,\sigma_j^2 
\;\sim\; 
\mathcal{N}\!\big(0,\ \sigma_j^2 \tau_j^2 \lambda_{jk}^2\big),
		\end{equation*}
		with local shrinkage $\lambda_{jk}^2$ and global shrinkage $\tau_j^2$. Following \citet{carvalho2010horseshoe}, the prior hierarchy is
\begin{align*}
			\lambda_{jk}^2 \mid \nu_{jk} &\sim \text{Inv-Gamma}(1,\, \nu_{jk}^{-1}), 
			& \nu_{jk} &\sim \text{Inv-Gamma}(1,\, 1), \\
			\tau_j^2 \mid \xi_j &\sim \text{Inv-Gamma}\!\left(\tfrac{m+1}{2},\ \xi_j^{-1}\right), 
			& \xi_j &\sim \text{Inv-Gamma}(1,\, 1),
		\end{align*}
		where $m = p-1$. 
The residual variance is assigned an inverse-gamma prior, $ \;
			\sigma_j^2 \sim \text{Inv-Gamma}(a_0, b_0).$ Note that, here we use the shape--rate parameterization, with density of Inv-Gamma$(a,b)$ proportional to $x^{-a-1}\exp(-b/x)$.
		
		
		Conditional on $(\theta_j,\sigma_j^2)$, the latent $Z_{ij}$ has conditional mean $\mu_{ij} = z_{i,-j}^\top \theta_j$ and variance $\sigma_j^2$. The observation mechanism is defined as
		\begin{itemize}
			\item[(a)] If $Y_{ij}>c_j$ (uncensored), then $Z_{ij}=Y_{ij}$ is fully observed.
			\item[(b)] If $Y_{ij}=c_j$ (censored), then $Z_{ij}$ follows a truncated normal distribution,
			\begin{equation*}
				Z_{ij} \,\big|\, \mathcal{C}_{ij}=1 \ \sim\ \mathcal{N}(\mu_{ij},\sigma_j^2)\ \text{ with support } (-\infty, c_j].
			\end{equation*}
		\end{itemize}
		
\subsubsection{Posterior computation}
		
Posterior inference is carried out by means of a Gibbs sampler in which each block of parameters and latent variables is updated in turn. The sampler alternates between imputing censored latent values, updating regression parameters, adapting local and global shrinkage scales, and reconstructing the precision matrix. We now describe each component of the algorithm in detail.
	
	The first step of the sampler updates the latent Gaussian data $Z$. When an observation is uncensored, so that $Y_{ij}>c_j$, the corresponding latent value is deterministically set to $Z_{ij}=Y_{ij}$. For censored observations with $Y_{ij}=c_j$, the latent entry is drawn from a truncated normal distribution with mean $\mu_{ij}=z_{i,-j}^\top\theta_j$ and variance $\sigma_j^2$, truncated above at $c_j$:
\begin{equation}
    \label{eq_latent_for_left_censoring}
 	Z_{ij} \;\sim\; \mathcal{N}(\mu_{ij},\sigma_j^2)\;\;\text{restricted to }(-\infty,c_j].   
\end{equation}

Given the augmented data, the regression coefficients $\theta_j$ are updated from their Gaussian full conditional. For each node $j \in \{1,\dots,p\}$, define the response vector $y_j = Z_{\cdot,j}$ and the predictor matrix $X^{(j)} = Z_{\cdot,-j}$. The conditional Gaussian likelihood yields the regression model
\[
y_j = X^{(j)} \theta_j + \varepsilon_j, \qquad \varepsilon_j \sim \mathcal{N}(0, \sigma_j^2 I_n).
\]
The horseshoe prior is placed on the regression coefficients:
\[
\theta_{jk} \mid \lambda_{jk}, \tau_j, \sigma_j^2 \sim \mathcal{N}(0, \sigma_j^2 \tau_j^2 \lambda_{jk}^2), 
\quad 
\lambda_{jk} \sim \mathcal{C}^+(0,1), 
\quad 
\tau_j \sim \mathcal{C}^+(0,1).
\]

The posterior updates proceed as follows. The conditional covariance and mean of the regression coefficients are
\[
V_j = \left( X^{(j)^\top} X^{(j)} + \operatorname{diag}\!\left(\frac{1}{\tau_j^2 \lambda_{j\cdot}^2}\right) \right)^{-1}, 
\qquad
\mu_{\theta_j} = V_j X^{(j)\top} y_j,
\]
and the coefficients are updated via $\theta_j \sim \mathcal{N}(\mu_{\theta_j}, \sigma_j^2 V_j).$

The residual variance is updated from its inverse-gamma conditional posterior,
\[
\sigma_j^2 \sim \operatorname{Inv\text{-}Gamma}\!\left( a_0 + \frac{n}{2}, \;
b_0 + \frac{1}{2}\|y_j - X^{(j)}\theta_j\|^2 \right).
\]
The local scales are updated using the Makalic–Schmidt parameterization,
\[
\lambda_{jk}^2 \sim \operatorname{Inv\text{-}Gamma}\!\left( 1, \; \frac{1}{\nu_{jk}} + \frac{\theta_{jk}^2}{2 \sigma_j^2 \tau_j^2} \right),
\qquad
\nu_{jk} \sim \operatorname{Inv\text{-}Gamma}(1, 1 + 1/\lambda_{jk}^2),
\]
and the global scale parameter is updated as
\[
\tau_j^2 \sim \operatorname{Inv\text{-}Gamma}\!\left( 
\frac{m+1}{2}, \; \frac{1}{\xi_j} + \sum_k \frac{\theta_{jk}^2}{2\sigma_j^2 \lambda_{jk}^2} \right),
\qquad
\xi_j \sim \operatorname{Inv\text{-}Gamma}(1, 1 + 1/\tau_j^2).
\]

Once all nodewise regressions are updated, the precision matrix $\Omega$ is reconstructed. The diagonal elements are determined by the inverse residual variances: $\Omega_{jj} = 1/\sigma_j^2,$
and the off-diagonal elements are obtained from the regression coefficients: $\Omega_{-j,j} = -\theta_j/\sigma_j^2.$
Because each regression is updated separately, the reconstructed matrix may not be perfectly symmetric. A symmetrization step is therefore applied, for example by averaging upper and lower triangular entries. If necessary, the resulting matrix can be projected to the cone of symmetric positive definite matrices to ensure validity.

After discarding an initial burn-in period, posterior summaries such as the mean and median precision matrices are computed as
\[
\hat{\Omega}_{\text{mean}} = \frac{1}{n_{\text{keep}}} \sum_{t \in \text{kept}} \Omega^{(t)}, 
\qquad
\hat{\Omega}_{\text{median}} = \operatorname{median}_{t \in \text{kept}} \Omega^{(t)}.
\]
	
\subsubsection{Initialization and practical considerations}
		
Latent $Z$ is initialized by setting $Z_{ij}=Y_{ij}$ when uncensored, and $Z_{ij}=c_j-\epsilon_{ij}$ for censored entries with $\epsilon_{ij}\sim | \mathcal{N} (0,1)|$. Regression parameters are initialized at neutral values ($\theta_j=0$, $\sigma_j^2=1$, $\lambda_{jk}^2=\nu_{jk}=\tau_j^2=\xi_j=1$).  Default hyperparameters $(a_0,b_0)=(10^{-2},10^{-2})$ provide weakly informative priors for variance terms, while the hierarchical horseshoe prior adaptively shrinks small coefficients without overshrinking large signals.
Efficient truncated normal sampling relies on stable implementations of the inverse-cdf method, with exponential tilting for extreme truncation. Linear algebra computations use Cholesky factorization of $A$, with Gaussian sampling via $\theta_j = \mu_\theta + \sigma_j\,L^{-\top} U$ where $LL^\top=A$ and $U \sim \mathcal{N}(0,I)$.  
				
\subsection{Extension to right-censored data}	

Extension to right-censored data is similar, that is rather focus on $ (-\infty,c] $, we use $ [c, \infty) $.
 Formally, the conditional distribution takes the form
\[
Z_{ij} \mid \widetilde{Y}_{ij}, \Omega, Z_{i,-j} \sim
\begin{cases}
	\mathcal{N}(\mu_{ij}, \sigma_j^2), & \text{if } \widetilde{Y}_{ij} = Z_{ij}, \\
	\mathcal{N}(\mu_{ij}, \sigma_j^2) \, \mathbb{I}(Z_{ij} \geq c_j), & \text{if } \widetilde{Y}_{ij} = c_j,
\end{cases}
\]
Thus, \eqref{eq_latent_for_left_censoring} is replaced by  
\begin{equation}
    \label{eq_latent_for_right_censoring}
 	Z_{ij} \;\sim\; \mathcal{N}(\mu_{ij},\sigma_j^2)\;\;\text{restricted to } [ c_j, \infty).   
\end{equation}
and the other step of the algorithm remain unchanged.

	\section{Graphical Horseshoe with Missing Data}
\label{sc_missing}	
	\subsection{Model setup and prior specification}
	
	Consider an i.i.d. sample $Y = (Y_1, \dots, Y_n) \in \mathbb{R}^{n \times p}$ from a $p$-dimensional Gaussian distribution with mean zero and precision matrix $\Omega \succ 0$. In practice, many data sets of this type contain missing values. Let $\mathcal{O} \subseteq \{1, \dots, n\} \times \{1, \dots, p\}$ denote the set of observed entries and $\mathcal{M} = \mathcal{O}^c$ the set of missing entries. For $(i,j) \in \mathcal{O}$ we observe $\widetilde{Y}_{ij}$, while values in $\mathcal{M}$ remain unobserved. The inferential goal is to recover the sparse precision matrix $\Omega$ under such incomplete data.
	
	Throughout, we assume the data are missing at random, in the sense that the probability of missingness depends only on observed variables and not on the unobserved values themselves. Under this assumption, the missingness mechanism can be ignored for likelihood-based inference, which allows us to formulate a Bayesian estimation procedure through data augmentation.
	
	The observed-data likelihood is obtained by marginalizing over the missing components of each $Y_i$. Writing $\mathcal{O}_i$ for the set of observed coordinates in row $i$, the distribution of $\widetilde{Y}_i$ is multivariate Gaussian with mean zero and covariance matrix given by the corresponding principal submatrix of $\Sigma = \Omega^{-1}$. Hence,
\begin{equation*}
  	L(\Omega; \{\widetilde{Y}_i\}_{i=1}^n) = \prod_{i=1}^n p(\widetilde{Y}_i \mid \Omega),  
\end{equation*}
	with
$
	\widetilde{Y}_i \sim \mathcal{N}_{|\mathcal{O}_i|}\bigl(0, \Sigma_{\mathcal{O}_i, \mathcal{O}_i}\bigr).
$
	Equivalently, the likelihood admits the form
\begin{equation}
\label{eq_likelihood_missing}
 	L(\Omega; \widetilde{Y}) = \prod_{i=1}^n (2\pi)^{-|\mathcal{O}_i|/2} 
	\left| \Sigma_{\mathcal{O}_i, \mathcal{O}_i} \right|^{-1/2} 
	\exp\!\left( -\frac{1}{2} \widetilde{Y}_i^\top \Sigma_{\mathcal{O}_i, \mathcal{O}_i}^{-1} \widetilde{Y}_i \right),   
\end{equation}
	where $\Sigma = \Omega^{-1}$. This representation emphasizes that only the observed entries contribute to the likelihood, while the unobserved entries are integrated out.
	
Here, we use the same graphical Horseshoe prior as in the previous section, given in \eqref{eq_horseshoe_prior}.
	
	\subsection{Posterior computation via Gibbs sampling}

Posterior inference under the graphical Horseshoe prior with missing data is performed via a Gibbs sampling algorithm. The sampler alternates between imputing missing entries of the latent Gaussian matrix and updating the precision matrix under the graphical Horseshoe prior.

\subsubsection{Initialization}

The Gibbs sampler begins with initial values for both the latent data matrix $Z$ and the precision matrix $\Omega$. For the missing entries, simple imputations such as column means or draws from a Gaussian fitted to the observed components of each variable provide reasonable starting values. The initial matrix is therefore
\[
Z^{(0)}_{ij} =
\begin{cases}
	\widetilde{Y}_{ij}, & (i,j) \in \mathcal{O}, \\
	\text{imputed value}, & (i,j) \in \mathcal{M}.
\end{cases}
\]
The precision matrix $\Omega$ must be initialized as a symmetric positive definite matrix. The identity matrix is the simplest choice, while a ridge-regularized inverse covariance estimate can yield faster convergence. In all cases, the initialization must satisfy $\Omega^{(0)} \succ 0$.

\subsubsection{Imputation of missing values}

At each iteration, missing entries in $Z$ are resampled from their conditional Gaussian distributions given the observed components and the current covariance estimate. Let $\Sigma^{(t-1)} = (\Omega^{(t-1)})^{-1}$ and partition it into blocks corresponding to observed indices $o$ and missing indices $m$ for a given row $i$. The conditional mean and covariance of the missing values are
\[
\mu_{m} = \Sigma_{mo} \Sigma_{oo}^{-1} Z_{i,o}, 
\qquad
\Sigma_{m} = \Sigma_{mm} - \Sigma_{mo} \Sigma_{oo}^{-1} \Sigma_{om}.
\]
The missing entries are then imputed as
\[
Z_{i,m} \sim \mathcal{N}(\mu_{m}, \Sigma_{m}).
\]

\subsubsection{Nodewise regression updates}

With the full latent data matrix $Z$ in hand, the precision matrix is updated through nodewise regressions under the graphical Horseshoe prior, as in Section~\ref{sc_nodewide_regression}. We omit repeating the same here for the sake of brevity.

The completed latent data matrix $Z$ is simultaneously available, offering a coherent Bayesian framework that unifies imputation and sparse graphical modeling in high-dimensional Gaussian settings.

\section{Numerical studies}
\label{sc_numerical}

Our methods for both cases of censored and missing data are implemented in the \texttt{R} package \texttt{GHScenmis} available on Github: \url{https://github.com/tienmt/ghscenmis}.

\subsection{Simulation setup and compared methods}

We compare our proposed methods against the Lasso approach tailored for censored and missing data as in \cite{augugliaro2020penalized, stadler2012missing}, with software implementations available through the \texttt{R} package \texttt{cglasso}  \citep{augugliaro2023cglasso}. Our Gibbs samplers are run with 5000 samples, with initial 1000 samples as burn-in. The \texttt{cglasso} method is run with default options and its tuning parameter is selected via AIC information criteria. 

We consider two settings for generating the true precision matrix ($\Omega_{\text{true}}$) and the corresponding covariance matrix ($\Sigma_{\text{true}}$) in our simulation study.

\paragraph{Setting 1:} Tridiagonal Precision Matrix.
\\
In the first scenario, we construct a tridiagonal precision matrix with off-diagonal elements representing weak conditional dependence between adjacent variables. Formally, let $p$ denote the number of variables, and define a $p \times p$ matrix $A$ with
$$
A_{ij} = 
\begin{cases} 
0.3, & \text{if } |i-j|=1 \\ 
0, & \text{otherwise}
\end{cases}
$$
The true precision matrix is then
$
\Omega_{\text{true}} = \text{diag}(1, \dots, 1) + A
$.
The corresponding covariance matrix is obtained by matrix inversion:
$
\;
\Sigma_{\text{true}} = \Omega_{\text{true}}^{-1}.
$
This structure mimics a simple chain graph where each variable depends on its immediate neighbors.

\paragraph{Setting 2:} Small Fully Connected Subset.
\\
In the second scenario, we introduce a small, densely connected subset of variables to study the effect of moderate correlations among a few variables. We define the covariance matrix $\Sigma_{\text{true}}$ as the identity matrix with added off-diagonal entries of 0.5 among the first three variables:
$$
\Sigma_{\text{true}} = 
\begin{pmatrix}
1 & 0.5 & 0.5 & 0 & \cdots & 0 \\
0.5 & 1 & 0.5 & 0 & \cdots & 0 \\
0.5 & 0.5 & 1 & 0 & \cdots & 0 \\
0 & 0 & 0 & 1 & & 0 \\
\vdots & \vdots & \vdots & & \ddots & \vdots \\
0 & 0 & 0 & 0 & \cdots & 1
\end{pmatrix}.
$$
The corresponding precision matrix is then computed as
$
\Omega_{\text{true}} = \Sigma_{\text{true}}^{-1}
,
$ which is still a very sparse matrix.
This setting allows us to examine estimation performance in the presence of a small but strong correlation block, contrasting with the sparse tridiagonal structure in Setting 1.

We evaluate each estimate from a method, $ \widehat{\Omega} $,  by considering squared $\ell_2$ norm error, given as $
\| \widehat{\Omega} - \Omega_{\rm true} \|_2^2.$ Each simulation setup is repeated 100 times.

\subsection{Simulations results with censored data}
	
We begin by presenting simulation studies in which the data are censored under a wide variety of scenarios.

\paragraph{Results with increasing dimensions.}
We set the sample size to $n=200$ and vary the dimension $p \in \{10, 20, 30\}$. Data are generated from a multivariate Gaussian distribution with the true precision matrix corresponding to either Setting I or Setting II (as described above). The generated data are then left-censored at the fixed threshold $c = (-0.5, 0.5, -0.5, 0.5, \ldots, -0.5, 0.5)$. This procedure is repeated 100 times, and the resulting estimation errors are summarized using  in Table \ref{tb_censored_dim_p}.

The results show a clear trend: as the dimension increases, both HScen and the censored graphical Lasso (cglasso) exhibit larger errors. Nevertheless, our Bayesian method consistently achieves smaller errors, highlighting its advantage in high-dimensional censored settings.

\begin{table}[!ht]
\centering
\caption{Simulation results with increasing dimensions (the number of variables) for censored data with $  n = 200 $, with censored vector is fixed at $ c = (-0.5 , 0.5, \ldots, -0.5 , 0.5 ) $. TPR: true positive rate and FDR: false positive rate. }
\begin{tabular}{ l  | ccc | ccc }
		\hline \hline
 Method
& $ \| \widehat{\beta} -\beta_0 \|_2^2 $ 
& TPR & FDR
& $ \| \widehat{\beta} -\beta_0 \|_2^2 $ 
& TPR & FDR
\\
\hline
& \multicolumn{3}{ c }{ Setting I, $ p =10 $ }
& \multicolumn{3}{ c }{ Setting II, $ p =10 $ }
\\
\hline
 cgLasso
& 2.13 (0.21) & 0.00 (0.01) & 0.00 (0.00) 
& 1.91 (2.02) & 1.00 (0.00) & 0.59 (0.40) 
\\
 cenGHS
& 0.65 (0.23) & 0.82 (0.11) & 0.01 (0.02)
& 0.65 (0.38) & 0.92 (0.15) & 0.00 (0.01) 
\\
\hline
& \multicolumn{3}{ c }{ Setting I, $ p = 20 $ }
& \multicolumn{3}{ c }{ Setting II, $ p = 20 $ }
\\
\hline
 cgLasso
& 4.44 (0.19) & 0.00 (0.01) & 0.00 (0.00)  
& 2.44 (2.47) & 1.00 (0.03) & 0.27 (0.35)  
\\
 cenGHS
& 1.60 (0.42) & 0.71 (0.10) & 0.01 (0.01) 
& 0.96 (0.37) & 0.87 (0.16) & 0.00 (0.00) 
\\
\hline
& \multicolumn{3}{ c }{ Setting I, $ p = 30 $ }
& \multicolumn{3}{ c }{ Setting II, $ p = 30 $ }
\\
\hline
 cgLasso
& 6.69 (0.26) & 0.00 (0.01) & 0.00 (0.00)  
& 3.04 (4.63) & 1.00 (0.03) & 0.11 (0.26)  
\\
 cenGHS
& 2.53 (0.44) & 0.61 (0.10) & 0.00 (0.00) 
& 1.42 (0.75) & 0.79 (0.16) & 0.00 (0.00) 
\\
\hline
\end{tabular}
\label{tb_censored_dim_p}
\end{table}

\paragraph{Results with increasing proportion of data being censored.}
We next consider the case with fixed sample size $n=200$ and dimension $p=10$. Instead of applying censoring at a fixed vector, we examine the impact of censoring at different proportions: $10\%$, $20\%$, and $30\%$ of the data. Specifically, data are generated from a multivariate Gaussian distribution with the true precision matrix given by either Setting I or Setting II (as described above), and censoring is applied at the empirical quantiles corresponding to the desired proportions. The procedure is repeated 100 times, and the resulting estimation errors are summarized in Table \ref{tb_censored_percent}.

The results indicate that both HScen and cglasso are affected by the degree of censoring. Nevertheless, our Bayesian method HScen consistently outperforms cglasso across all scenarios, with the performance gap becoming particularly pronounced when $50\%$ of the data are censored.

\begin{table}[!ht]
\centering
\caption{Simulation results for censored data with $  n = 200, p = 10 $, with increasing the amount of data being censored from 10\%, 20\%, and 30\% . TPR: true positive rate and FDR: false positive rate. }
\begin{tabular}{ l  | ccc | ccc }
		\hline \hline
 Method
& $ \| \widehat{\beta} -\beta_0 \|_2^2 $ 
& TPR & FDR
& $ \| \widehat{\beta} -\beta_0 \|_2^2 $ 
& TPR & FDR
\\
\hline
& \multicolumn{3}{ c }{ Setting I, 10\% censored }
& \multicolumn{3}{ c }{ Setting II, 10\% censored }
\\
\hline
 cgLasso
& 0.45 (0.34) & 1.00 (0.00) & 0.51 (0.24)
& 0.41 (0.19) & 1.00 (0.00) & 0.28 (0.15)
\\
 cenGHS
& 0.38 (0.13) & 0.99 (0.04) & 0.03 (0.03)
& 0.39 (0.18) & 0.99 (0.05) & 0.01 (0.02) 
\\
\hline
& \multicolumn{3}{ c }{ Setting I, 20\% censored }
& \multicolumn{3}{ c }{ Setting II, 20\% censored }
\\
\hline
 cgLasso
& 0.54 (0.26) & 1.00 (0.00) & 0.62 (0.29) 
& 0.55 (0.41) & 1.00 (0.00) & 0.43 (0.29)  
\\
 cenGHS
& 0.41 (0.14) & 0.98 (0.04) & 0.04 (0.03)  
& 0.42 (0.18) & 1.00 (0.03) & 0.01 (0.02) 
\\
\hline
& \multicolumn{3}{ c }{ Setting I, 30\% censored }
& \multicolumn{3}{ c }{ Setting II, 30\% censored }
\\
\hline
 cgLasso
& 0.66 (0.22) & 0.99 (0.03) & 0.54 (0.35)  
& 0.87 (0.60) & 1.00 (0.00) & 0.62 (0.35)
\\
 cenGHS
& 0.46 (0.14) & 0.96 (0.06) & 0.03 (0.03) 
& 0.46 (0.23) & 0.97 (0.09) & 0.01 (0.02) 
\\
\hline
\end{tabular}
\label{tb_censored_percent}
\end{table}

\paragraph{Results with increasing sample size.}
We next investigate the effect of increasing the sample size on estimation accuracy. The dimension is fixed at $p=10$, and censoring is applied at the fixed vector $(-0.5, 0.5, -0.5, 0.5, \ldots, -0.5, 0.5)$. Sample sizes are varied across $n = 200, 500, 1000$. Data are generated from a multivariate Gaussian distribution with the true precision matrix corresponding to either Setting I or Setting II (as described above). Results over 100 replications are summarized in Table \ref{tb_censored_sample_n}.

As expected, both HScen and cglasso show reduced estimation error as the sample size increases. However, the Bayesian method HScen demonstrates a clear advantage, achieving substantially smaller errors, particularly at larger sample sizes. Moreover, cglasso performs poorly under Setting I, where the underlying graph structure follows a chain, further highlighting the robustness of our approach.

\begin{table}[!ht]
\centering
\caption{Simulation results with increasing the sample sizes for censored data with $  p = 10 $, with censored vector is fixed at $ c = (-0.5 , 0.5, \ldots, -0.5 , 0.5 ) $. TPR: true positive rate and FDR: false positive rate. }
\begin{tabular}{ l  | ccc | ccc }
		\hline \hline
 Method
& $ \| \widehat{\beta} -\beta_0 \|_2^2 $ 
& TPR & FDR
& $ \| \widehat{\beta} -\beta_0 \|_2^2 $ 
& TPR & FDR
\\
\hline
& \multicolumn{3}{ c }{ Setting I, $ n = 200 $ }
& \multicolumn{3}{ c }{ Setting II, $ n = 200 $ }
\\
\hline
 cgLasso
& 2.13 (0.21) & 0.00 (0.01) & 0.00 (0.00) 
& 1.91 (2.02) & 1.00 (0.00) & 0.59 (0.40) 
\\
 cenGHS
& 0.65 (0.23) & 0.82 (0.11) & 0.01 (0.02)
& 0.65 (0.38) & 0.92 (0.15) & 0.00 (0.01) 
\\
\hline
& \multicolumn{3}{ c }{ Setting I, $ n = 500 $ }
& \multicolumn{3}{ c }{ Setting II, $ n = 500 $ }
\\
\hline
 cgLasso
& 2.01 (0.09) & 0.00 (0.00) & 0.00 (0.00)  
& 1.00 (0.63) & 1.00 (0.00) & 0.72 (0.41)  
\\
 cenGHS
& 0.24 (0.08) & 1.00 (0.00) & 0.02 (0.02) 
& 0.21 (0.10) & 1.00 (0.00) & 0.01 (0.01) 
\\
\hline
& \multicolumn{3}{ c }{ Setting I, $ n = 1000 $ }
& \multicolumn{3}{ c }{ Setting II, $ n = 1000 $ }
\\
\hline
 cgLasso
& 1.97 (0.08) & 0.04 (0.15) & 0.00 (0.00)  
& 0.82 (0.51) & 1.00 (0.00) & 0.76 (0.41)  
\\
 cenGHS
& 0.11 (0.03) & 1.00 (0.00) & 0.02 (0.02) 
& 0.10 (0.05) & 1.00 (0.00) & 0.01 (0.01) 
\\
\hline
\end{tabular}
\label{tb_censored_sample_n}
\end{table}

\subsection{Simulations with missing data}	

\paragraph{Results with increasing dimensions.}  
We fix the sample size at $n=200$ and vary the dimension $p \in \{10, 20, 30\}$. Data are generated from a multivariate Gaussian distribution with the true precision matrix specified by either Setting I or Setting II (as described above). To mimic missingness, $10\%$ of the simulated observations are removed uniformly at random. This procedure is repeated 100 times, and the resulting estimation errors are displayed as  in Table \ref{tb_missing_dimension_p}.

The results show that the estimation errors of both methods increase with dimension. Nonetheless, the Bayesian method HScen consistently achieves smaller errors than cglasso, with the performance gap being most pronounced when the true precision matrix corresponds to Setting II.

\begin{table}[!ht]
\centering
\caption{Simulation results for 10\% missing data with $  n = 200 $ and increase dimension. TPR: true positive rate and FDR: false discovery rate. }
\begin{tabular}{ l  | ccc | ccc }
		\hline \hline
 Method
& $ \| \widehat{\beta} -\beta_0 \|_2^2 $ 
& TPR & FDR
& $ \| \widehat{\beta} -\beta_0 \|_2^2 $ 
& TPR & FDR
\\
\hline
& \multicolumn{3}{ c }{ Setting I, $ p =10 $ }
& \multicolumn{3}{ c }{ Setting II, $ p =10 $ }
\\
\hline
 cgLasso
& 0.62 (0.37) & 1.00 (0.00) & 0.66 (0.30) 
& 0.45 (0.30) & 1.00 (0.00) & 0.34 (0.20) 
\\
 cenGHS
& 0.38 (0.12) & 0.96 (0.06) & 0.02 (0.03) 
& 0.37 (0.17) & 0.97 (0.09) & 0.01 (0.01)  
\\
\hline
& \multicolumn{3}{ c }{ Setting I, $ p = 20 $ }
& \multicolumn{3}{ c }{ Setting II, $ p = 20 $ }
\\
\hline
 cgLasso
& 0.93 (0.14) & 1.00 (0.00) & 0.42 (0.07)  
& 0.71 (0.20) & 1.00 (0.00) & 0.09 (0.09)  
\\
 cenGHS
& 0.91 (0.18) & 0.90 (0.07) & 0.01 (0.01) 
& 0.58 (0.23) & 0.95 (0.12) & 0.00 (0.00) 
\\
\hline
& \multicolumn{3}{ c }{ Setting I, $ p = 30 $ }
& \multicolumn{3}{ c }{ Setting II, $ p = 30 $ }
\\
\hline
 cgLasso
& 1.66 (0.20) & 1.00 (0.01) & 0.33 (0.12)  
& 0.98 (0.21) & 1.00 (0.00) & 0.03 (0.04) 
\\
 cenGHS
& 1.42 (0.25) & 0.85 (0.07) & 0.00 (0.00) 
& 0.78 (0.31) & 0.92 (0.15) & 0.00 (0.00) 
\\
\hline
\end{tabular}
\label{tb_missing_dimension_p}
\end{table}

\paragraph{Results with increasing sample size.}
We next examine how increasing the sample size affects estimation accuracy. The dimension is fixed at $p=10$, and sample sizes are set to $n = 200, 500, 1000$. Data are generated from a multivariate Gaussian distribution with the true precision matrix corresponding to either Setting I or Setting II (as described above), after which $10\%$ of the data are removed uniformly at random. Results over 100 replications are presented in Tables \ref{tb_missing_sample_n}.

As expected, both methods achieve lower errors as the sample size increases. However, the Bayesian approach HScen consistently outperforms cglasso across all scenarios, demonstrating superior robustness to missing data.

\begin{table}[!ht]
\centering
\caption{Simulation results for 10\% missing data with $ p = 10 $ increase sample size. TPR: true positive rate and FDR: false discovery rate. }
\begin{tabular}{ l  | ccc | ccc }
		\hline \hline
 Method
& $ \| \widehat{\beta} -\beta_0 \|_2^2 $ 
& TPR & FDR
& $ \| \widehat{\beta} -\beta_0 \|_2^2 $ 
& TPR & FDR
\\
\hline
& \multicolumn{3}{ c }{ Setting I, $ n = 200 $ }
& \multicolumn{3}{ c }{ Setting II, $ n = 200 $ }
\\
\hline
 cgLasso
& 0.62 (0.37) & 1.00 (0.00) & 0.66 (0.30) 
& 0.45 (0.30) & 1.00 (0.00) & 0.34 (0.20) 
\\
 cenGHS
& 0.38 (0.12) & 0.96 (0.06) & 0.02 (0.03) 
& 0.37 (0.17) & 0.97 (0.09) & 0.01 (0.01)  
\\
\hline
& \multicolumn{3}{ c }{ Setting I, $ n = 500 $ }
& \multicolumn{3}{ c }{ Setting II, $ n = 500 $ }
\\
\hline
 cgLasso
& 0.27 (0.08) & 1.00 (0.00) & 0.86 (0.29)   
&   0.19 (0.12) & 1.00 (0.00) & 0.28 (0.24) 
\\
 cenGHS
& 0.14 (0.05) & 1.00 (0.00) & 0.03 (0.03)
& 0.14 (0.06) & 1.00 (0.00) & 0.01 (0.02)  
\\
\hline
& \multicolumn{3}{ c }{ Setting I, $ n = 1000 $ }
& \multicolumn{3}{ c }{ Setting II, $ n = 1000 $ }
\\
\hline
 cgLasso
& 0.13 (0.03) & 1.00 (0.00) & 1.00 (0.00)  
& 0.17 (0.05) & 1.00 (0.00) & 0.58 (0.46) 
\\
 cenGHS
& 0.07 (0.02) & 1.00 (0.00) & 0.03 (0.03) 
& 0.07 (0.03) & 1.00 (0.00) & 0.01 (0.02) 
\\
\hline
\end{tabular}
\label{tb_missing_sample_n}
\end{table}

\paragraph{Results with increasing proportion of missing data.}
We now fix the sample size at $n=200$ and dimension at $p=10$, and investigate the effect of varying the proportion of missing data. Specifically, we consider missingness levels of $10\%$, $20\%$, and $30\%$. Data are generated from a multivariate Gaussian distribution with the true precision matrix corresponding to either Setting I or Setting II (as described above). The specified proportion of entries is then removed uniformly at random. This process is repeated 100 times, and the resulting estimation errors are displayed as  in Table \ref{tb_mis_missing_percent}.

As expected, both methods exhibit a modest increase in estimation error as the proportion of missing data rises. Nevertheless, the Bayesian method cenGHS consistently outperforms cglasso, yielding smaller and more stable errors across all levels of missingness.

Some trace and acf plots are given in Appendix~\ref{app:addl_simu_results} together with effective sample size of the MCMC samples to demonstrate the behavior of the algorithm.

\begin{table}[!ht]
\centering
\caption{Simulation results for censored data with $  n = 200, p = 20 $, with increasing the amount of data being censored from 10\%, 20\%, and 30\% . TPR: true positive rate and FDR: false positive rate. }
\begin{tabular}{ l  | ccc | ccc }
		\hline \hline
 Method
& $ \| \widehat{\beta} -\beta_0 \|_2^2 $ 
& TPR & FDR
& $ \| \widehat{\beta} -\beta_0 \|_2^2 $ 
& TPR & FDR
\\
\hline
& \multicolumn{3}{ c }{ Setting I, missing 10\%  }
& \multicolumn{3}{ c }{ Setting II, missing 10\%  }
\\
\hline
 cgLasso
& 0.93 (0.14) & 1.00 (0.00) & 0.42 (0.07)  
& 0.71 (0.20) & 1.00 (0.00) & 0.09 (0.09)  
\\
 cenGHS
& 0.91 (0.18) & 0.90 (0.07) & 0.01 (0.01) 
& 0.58 (0.23) & 0.95 (0.12) & 0.00 (0.00) 
\\
\hline
& \multicolumn{3}{ c }{ Setting I, missing 20\%  }
& \multicolumn{3}{ c }{ Setting II, missing 20\%  }
\\
\hline
 cgLasso
& 1.65 (1.24) & 1.00 (0.01) & 0.53 (0.22) 
& 0.89 (0.22) & 1.00 (0.00) & 0.08 (0.11)  
\\
 cenGHS
& 1.15 (0.21) & 0.75 (0.09) & 0.00 (0.00)  
& 0.67 (0.31) & 0.90 (0.15) & 0.00 (0.00) 
\\
\hline
& \multicolumn{3}{ c }{ Setting I, missing 30\% }
& \multicolumn{3}{ c }{ Setting II, missing 30\%  }
\\
\hline
 cgLasso
& 1.94 (0.78) & 0.99 (0.02) & 0.61 (0.26) 
& 1.17 (0.26) & 0.99 (0.06) & 0.05 (0.06) 
\\
 cenGHS
& 1.71 (0.26) & 0.49 (0.10) & 0.00 (0.00) 
& 0.81 (0.30) & 0.72 (0.16) & 0.00 (0.00) 
\\
\hline
\end{tabular}
\label{tb_mis_missing_percent}
\end{table}


\section{Application to single cell-data: megakaryocyte-erythroid progenitors}
\label{sc_real_Data}

In a study of blood cell formation, \citet{psaila2016single} identified three distinct sub-populations of cells derived from hematopoietic stem cells through differentiation. One of these sub-populations, termed MK-MEP, is a previously uncharacterized, rare population of bipotent cells that primarily give rise to megakaryocytic progeny. The data is available from the \texttt{cglasso} package

Multiplex RT-qPCR was used to profile the expression of 63 genes across 48 single human MK-MEP cells. RT-qPCR measurements are typically right-censored, with a manufacturer-specified detection limit of 40 (the censored level). Raw expression values were mean-normalized following the approach of \citet{pipelers2017unified}. Further details are provided in \citet{augugliaro2020penalized}.

\begin{figure}[!ht]
    \centering
    \includegraphics[width=14cm]{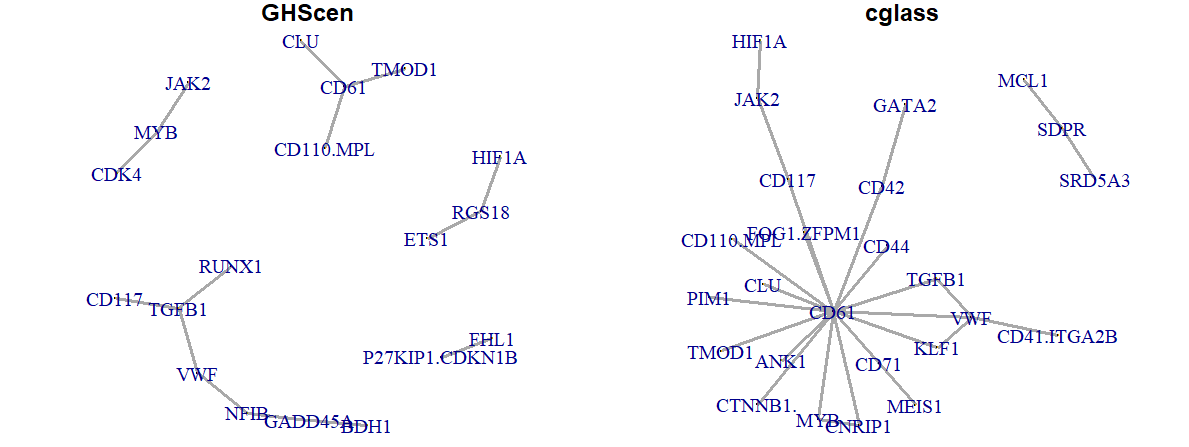}
    \caption{Graphs from different methods on the real data set MKMEP. We see that GHSsen return a  sparse networks compared to cglasso.}
    \label{fg:real_Data_graph}
\end{figure}

We applied the GHScen method for 10,000 iterations, discarding the first 1,000 iterations as burn-in. Figure \ref{fg:real_Data_graph} shows the estimated graphs obtained from different methods. For GHScen, nonzero entries in the precision matrix were identified using 95\% credible intervals that do not include zero. Compared with cglasso, the graph estimated by GHScen is notably sparser and reveals five distinct connected subgroups, whereas cglasso produces a single large connected component.

\section{Theoretical results}
\label{sc_theory}

In this section, we study posterior contraction of the tempered (o, fractional) posterior for the proposed censored Graphical Horseshoe in high-dimensional Gaussian graphical models with missing or censored observations. The main technical difficulty arises from the observed-data likelihood: missingness leads to marginal Gaussian likelihoods, while censoring introduces non-linear log-CDF terms that distort curvature and preclude global log-concavity. We show that, despite these complications, the global--local shrinkage structure of the graphical Horseshoe prior yields sharp posterior concentration under standard sparsity and spectral regularity conditions.

Our analysis is conducted in terms of R\'enyi divergence, which provides a natural information-theoretic loss for tempered posteriors and remains well suited to likelihoods with non-quadratic structure. The results demonstrate that the posterior concentrates around the true precision matrix at the same order as in fully observed Gaussian graphical models, up to constants depending on the censoring mechanism.

Let $ p_{\Omega} $ and $ P_{\Omega} $ denote the density and the distribution of a Gaussian $ \mathcal{N} (0, \Omega^{-1} ) $ random variable. Denoting the prior on $\Omega$ by $\pi_{HS}(\Omega)$, 
the tempered posterior \citep{bhattacharya2016bayesian, alquier2020concentration,mai2025concentration,mai2025properties} is given by
\[\pi_{n,\alpha}(\Omega) \propto L^{\alpha}(\Omega; \tilde{Y}) \pi_{HS}(\Omega).\] We consider the following assumptions regarding the truth.
	
	\begin{assume}[Bounded spectrum]
\label{assum_true_precision_Spectrum}
		The  precision matrix $ \Omega, \Omega_0 $ lie in the spectral band
\[\mathcal S=\{\Omega\succ0:\ 0 < m \le \lambda_{\min}(\Omega) \le \lambda_{\max}(\Omega)\le M<\infty\}.
\]
	\end{assume}

	\begin{assume}
\label{assum_true_sparsity}
The true precision matrix $\Omega_0 $ is sparse in the sense that
\[
		\Omega_0 \in  \mathcal{S}(s) 
         := \left\lbrace 
		\Omega \in \mathcal{S} : \!\! 
		\sum_{ 1\leq i < j \leq p}
		\mathbf{1}_{(\omega_{ij} \neq 0)} \leq s, \, s < n
		\right\rbrace.
\]
	\end{assume}

\begin{assume}[Non-degenerate censoring probability.]
   \label{assume_CDF_of_censored} 
There exists a finite constant $ K_1 = K_1(m,M,c)$ such that for every $\Omega_t \in \mathcal S$ and every matrix $A \in \mathbb{R}^{p \times p}$, the following quadratic form involving the expected Hessian is bounded:
$$
| {\rm trace} (A^\top \mathbb{E}_{\Omega_0}[\nabla^2 \ell(\Omega_t; \widetilde Y)]  \: A) | 
\le 
K_1 \|A\|_F^2.
$$
 
\end{assume}

Assumption~\ref{assum_true_precision_Spectrum} is standard in both frequentist and Bayesian literature for sparse precision matrix estimation, \citep{cai2016estimating,jankova2017honest,banerjee2015bayesian,sagar2024precision,zhang2022contraction}. 
The sparsity condition in Assumption~\ref{assum_true_sparsity} is common in high-dimensional settings \citep{bellec2018slope,gao2020general,banerjee2021bayesian}.
Assumption~\ref{assume_CDF_of_censored} is a non-trivial assumption for censored data due to the log-CDF term in the likelihood. Its role is to control the local curvature of the observed-data log-likelihood, which, due to censoring, contains Gaussian distribution function terms rather than purely quadratic forms. These terms destroy global log-concavity and can, in principle, lead to regions of near-flat likelihood. The assumption requires that the expected Hessian of the censored log-likelihood remains uniformly bounded in Frobenius norm over a spectral neighborhood of the truth. This condition ensures that the information loss induced by censoring is controlled and that the likelihood retains sufficient local identifiability to support posterior concentration. In particular, it rules out degenerate censoring configurations in which large portions of the data contribute negligible information about the precision matrix.

From a statistical perspective, Assumption~\ref{assume_CDF_of_censored} can be interpreted as an information non-degeneracy condition: it guarantees that censoring does not eliminate curvature in directions corresponding to nonzero precision matrix entries. Similar curvature or restricted eigenvalue conditions are implicitly required in frequentist analyses of censored graphical models, although they are rarely stated explicitly. Here, the assumption allows the Bayesian posterior to concentrate at the same rate as in the fully observed case, up to constants depending on the censoring thresholds.

We now present first a consistency result of the tempered posterior in $ p > n $ setting.

Given two probability measures, $P$ and $R$, that are both dominated by a shared measure $\nu$, and for any  $\alpha$ in $(0,1)$, the $\alpha$-Rényi divergence between $P$ and $R$ is given by the formula:
\begin{align*}
\mathcal{D}_{\alpha}(P,R) =
\frac{1}{\alpha-1} \log \int
\left( \frac{{\rm d}P }{{\rm d} \nu} \right)^\alpha
\left( \frac{{\rm d}R }{{\rm d} \nu} \right)^{1-\alpha} {\rm d} \nu.
\end{align*}
Furthermore, the Kullback-Leibler divergence is defined as $
\mathcal{K} (P,R) =
\int \log \left(\frac{{\rm d}P}{{\rm d}R} \right){\rm d}P $. This definition holds provided that $P$ is absolutely continuous with respect to $R$ (denoted $P \ll R$). If this condition is not met, the divergence is taken to be $+ \infty$. We now present the results on posterior contraction for the case of censored data, followed by the same for missing data.

\begin{theorem}[Posterior contraction for censored data]
    \label{thm_censoredGHS}
Consider the model formulation for censored data as in Section~\ref{sc_censored_GHS}. For any $\alpha\in(0,1)$, under Assumptions~\ref{assum_true_precision_Spectrum}, 
 \ref{assum_true_sparsity}, 
 \ref{assume_CDF_of_censored}, we have that
		$$
		\mathbb{E} \left[ \int \mathcal{D}_{\alpha}(P_{\Omega},P_{\Omega_0}) \pi_{n,\alpha}({\rm d}\Omega ) \right]
		\leq \frac{1+\alpha}{1-\alpha}\varepsilon_n.
		$$
with $ \varepsilon_n = K n^{-1}s\log p $, for some universal constant $K>0$ depending only on $m, M$ and the censored value $c $.    
\end{theorem}

\begin{theorem}[Posterior contraction for missing data]
    \label{thm_missing_GHS}
    Consider the model formulation for missing data as in Section~\ref{sc_missing}. For any $\alpha\in(0,1)$, under Assumptions~\ref{assum_true_precision_Spectrum}, 
 \ref{assum_true_sparsity}, we have that
		$$
		\mathbb{E} \left[ \int \mathcal{D}_{\alpha}(P_{\Omega},P_{\Omega_0}) \pi_{n,\alpha}({\rm d}\Omega ) \right]
		\leq \frac{1+\alpha}{1-\alpha}\varepsilon_n.
		$$
with $ \varepsilon_n = K n^{-1}s\log p  $, for some universal constant $K>0$ does not depending on $n,p,s$.
\end{theorem}

\paragraph{Discussion of the results.}
Theorems~\ref{thm_censoredGHS} and ~\ref{thm_missing_GHS} establish that, for both missing and censored data, the tempered posterior contracts at rate $\varepsilon_n \asymp n^{-1}s\log p,$ which coincides with the optimal rate for sparse precision matrix estimation under complete observation \citep{sagar2024precision}.
Thus, incomplete observation does not fundamentally alter the statistical difficulty of the problem.

The distinction between the two regimes lies in likelihood geometry. For missing data, contraction follows from standard Kullback--Leibler control after marginalization. For censored data, additional regularity is required to control the curvature of the log-likelihood induced by Gaussian CDF terms. 

Importantly, the results are adaptive: the contraction rate depends on the unknown sparsity level~$s$, with no tuning of the Horseshoe prior. This confirms that global--local shrinkage remains theoretically stable under censoring and missingness. To our knowledge, these are the first posterior contraction results for sparse precision matrix estimation in the presence of censoring; comparable guarantees are currently unavailable for penalized likelihood methods such as the censored graphical lasso.

\section{Discussion and conclusion}
\label{sc_conclusion}
This work extends Bayesian precision matrix estimation to settings involving censored and missing (high-dimensional) Gaussian data by introducing the Censored Graphical Horseshoe (CGHS) model. Building on the adaptive shrinkage properties of the graphical Horseshoe prior, our formulation incorporates a latent-variable representation that naturally handles censored observations as well as arbitrarily missing entries. This generalization substantially broadens the applicability of the Horseshoe framework, enabling principled inference in problems where existing graphical modeling tools are not directly equipped to handle incomplete or censored measurements.

A key contribution of our paper is the development of an efficient Gibbs sampling algorithm for posterior computation. By alternating between imputing latent data, updating regression coefficients and error variances, adapting local and global shrinkage scales, and reconstructing the precision matrix, the sampler provides a tractable route to full Bayesian inference in models that would otherwise be computationally prohibitive. 

Perhaps the most distinctive aspect of this work is the theoretical analysis. We derive posterior concentration rates for both censored-data and missing-data regimes, establishing that CGHS recovers the underlying precision structure at near-optimal rates under standard sparsity conditions. To the best of our knowledge, no analogous results exist for frequentist or Bayesian versions of the graphical lasso in these regimes. Our theory therefore fills an important gap in the literature, demonstrating that global-local shrinkage priors remain stable and adaptively informative even when the likelihood is distorted by censoring or incomplete observations.

There are several promising avenues for future work. One direction is to extend CGHS beyond the Gaussian setting—for example, to copula-based or semiparametric graphical models—which would broaden its applicability to non-normal data structures. Another is to develop scalable posterior approximation methods, such as variational inference or stochastic-gradient MCMC, to improve computational efficiency in ultra-high-dimensional domains like genomics and neuroimaging. Incorporating Missing Not at Random (MNAR) mechanisms into the latent-variable framework represents an additional challenge with substantial practical relevance, particularly in biomedical studies where the missingness may depend on unobserved values. Finally, establishing formal model-selection strategies or credible-graph procedures would enable principled uncertainty quantification for inferred network structures.

\subsubsection*{Acknowledgments}
The views, findings, and opinions presented in this work are exclusively those of the authors and do not reflect the official stance of the Norwegian Institute of Public Health or IIM Indore.

\subsubsection*{Conflicts of interest/Competing interests}
The authors declare no potential conflict of interests.

	\bibliographystyle{apalike}
	\bibliography{ref_cgHS}

@article{gan2019bayesian,
	title={Bayesian regularization for graphical models with unequal shrinkage},
	author={Gan, Lingrui and Narisetty, Naveen N and Liang, Feng},
	journal={Journal of the American Statistical Association},
	volume={114},
	number={527},
	pages={1218--1231},
	year={2019},
	publisher={Taylor \&amp; F\textbf{}rancis}
}

@article{mai2025handling,
  title={Handling bounded response in high dimensions: a Horseshoe prior Bayesian Beta regression approach},
  author={Mai, The Tien},
  journal={arXiv preprint arXiv:2505.22211},
  year={2025}
}

@article{mai2025sparse,
  title={A sparse PAC-Bayesian approach for high-dimensional quantile prediction},
  author={Mai, The Tien},
  journal={Statistics and Computing},
  volume={35},
  number={4},
  pages={93},
  year={2025},
  publisher={Springer}
}

@article{mai2025tobit,
  title={High-dimensional Bayesian Tobit regression for censored response with Horseshoe prior},
  author={Mai, The Tien},
  journal={arXiv preprint arXiv:2505.08288},
  year={2025}
}

@article{mai2025high,
  title={On high-dimensional classification by sparse generalized bayesian logistic regression},
  author={Mai, The Tien},
  journal={Statistical Papers},
  year={2025}
}

@article{mai2025adaptive,
  title={Adaptive posterior concentration rates for sparse high-dimensional linear regression with random design and unknown error variance},
  author={Mai, The Tien},
  journal={Journal of the Korean Statistical Society},
  pages={1--35},
  year={2025},
  publisher={Springer}
}

@article{mai2025properties,
  title={On properties of fractional posterior in generalized reduced-rank regression},
  author={Mai, The Tien},
  journal={Journal of Multivariate Analysis},
  pages={105481},
  year={2025},
  publisher={Elsevier}
}

@article{psaila2016single,
  title={Single-cell profiling of human megakaryocyte-erythroid progenitors identifies distinct megakaryocyte and erythroid differentiation pathways},
  author={Psaila, Bethan and Barkas, Nikolaos and Iskander, Deena and Roy, Anindita and Anderson, Stacie and Ashley, Neil and Caputo, Valentina S and Lichtenberg, Jens and Loaiza, Sandra and Bodine, David M and others},
  journal={Genome biology},
  volume={17},
  number={1},
  pages={83},
  year={2016},
  publisher={Springer}
}

@article{pesonen2015covariance,
  title={Covariance matrix estimation for left-censored data},
  author={Pesonen, Maiju and Pesonen, Henri and Nevalainen, Jaakko},
  journal={Computational Statistics \& Data Analysis},
  volume={92},
  pages={13--25},
  year={2015},
  publisher={Elsevier}
}

@article{pipelers2017unified,
  title={A unified censored normal regression model for qPCR differential gene expression analysis},
  author={Pipelers, Peter and Clement, Lieven and Vynck, Matthijs and Hellemans, Jan and Vandesompele, Jo and Thas, Olivier},
  journal={Plos one},
  volume={12},
  number={8},
  pages={e0182832},
  year={2017},
  publisher={Public Library of Science San Francisco, CA USA}
}

@article{mccall2014non,
  title={On non-detects in qPCR data},
  author={McCall, Matthew N and McMurray, Helene R and Land, Hartmut and Almudevar, Anthony},
  journal={Bioinformatics},
  volume={30},
  number={16},
  pages={2310--2316},
  year={2014},
  publisher={Oxford University Press}
}

@article{derveaux2010successful,
  title={How to do successful gene expression analysis using real-time PCR},
  author={Derveaux, Stefaan and Vandesompele, Jo and Hellemans, Jan},
  journal={Methods},
  volume={50},
  number={4},
  pages={227--230},
  year={2010},
  publisher={Elsevier}
}

@article{augugliaro2023cglasso,
	title={cglasso: An R package for conditional graphical lasso inference with censored and missing values},
	author={Augugliaro, Luigi and Sottile, Gianluca and Wit, Ernst C and Vinciotti, Veronica},
	journal={Journal of Statistical Software},
	volume={105},
	pages={1--58},
	year={2023}
}

@article{stadler2012missing,
	title={Missing values: sparse inverse covariance estimation and an extension to sparse regression},
	author={St{\"a}dler, Nicolas and B{\"u}hlmann, Peter},
	journal={Statistics and Computing},
	volume={22},
	number={1},
	pages={219--235},
	year={2012},
	publisher={Springer}
}

@article{augugliaro2020penalized,
	title={ $\ell_1$-penalized censored gaussian graphical model},
	author={Augugliaro, Luigi and Abbruzzo, Antonino and Vinciotti, Veronica},
	journal={Biostatistics},
	volume={21},
	number={2},
	pages={e1--e16},
	year={2020},
	publisher={Oxford University Press}
}

@article{lounici2014high,
	title={High-dimensional covariance matrix estimation with missing observations},
	author={Lounici, Karim},
	journal={Bernoulli},
	volume={20},
	number={3},
	pages={1029--1058},
	year={2014}
}

@article{mai2025concentration,
	title={Concentration properties of fractional posterior in 1-bit matrix completion},
	author={Mai, The Tien},
	journal={Machine Learning},
	volume={114},
	number={1},
	pages={7},
	year={2025},
	publisher={Springer}
}

@article{ryali2012estimation,
	title={Estimation of functional connectivity in fMRI data using stability selection-based sparse partial correlation with elastic net penalty},
	author={Ryali, Srikanth and Chen, Tianwen and Supekar, Kaustubh and Menon, Vinod},
	journal={NeuroImage},
	volume={59},
	number={4},
	pages={3852--3861},
	year={2012},
	publisher={Elsevier}
}

@article{pourahmadi2011covariance,
	title={Covariance Estimation: The GLM and Regularization Perspectives},
	author={Pourahmadi, Mohsen},
	journal={Statistical Science},
	volume={26},
	number={3},
	pages={369--387},
	year={2011}
}

@article{callot2021nodewise,
	title={A nodewise regression approach to estimating large portfolios},
	author={Callot, Laurent and Caner, Mehmet and {\"O}nder, A {\"O}zlem and Ula{\c{s}}an, Esra},
	journal={Journal of Business \& Economic Statistics},
	volume={39},
	number={2},
	pages={520--531},
	year={2021},
	publisher={Taylor \& Francis}
}

@article{van2014asymptotically,
	title={On asymptotically optimal confidence regions and tests for high-dimensional models},
	author={Van de Geer, Sara and B{\"u}hlmann, Peter and Ritov, Ya’acov and Dezeure, Ruben and others},
	journal={The Annals of Statistics},
	volume={42},
	number={3},
	pages={1166--1202},
	year={2014},
	publisher={Institute of Mathematical Statistics}
}

@book{lauritzen1996,
	author = {Lauritzen, Steffen L.},
	isbn = {0-19-852219-3},
	publisher = {Oxford University Press},
	title = {Graphical Models},
	year = 1996
}

@article{friedman2008sparse,
	title = {Sparse Inverse Covariance Estimation with the Graphical Lasso},
	volume = {9},
	number = {3},
	journal = {Biostatistics},
	author = {Friedman, Jerome and Hastie, Trevor and Tibshirani, Robert},
	year = {2008},
	pages = {432--441}
}

@article{mai2024concentration,
	title={{Concentration of a Sparse Bayesian Model With Horseshoe Prior in Estimating High-Dimensional Precision Matrix}},
	author={Mai, The Tien},
	journal={Stat},
	volume={13},
	number={4},
	pages={e70008},
	year={2024},
	publisher={Wiley Online Library}
}

@article{carvalho2010horseshoe,
	title={The horseshoe estimator for sparse signals},
	author={Carvalho, Carlos M and Polson, Nicholas G and Scott, James G},
	journal={Biometrika},
	pages={465--480},
	year={2010},
	publisher={JSTOR}
}

@article{fan2009network,
	title={Network exploration via the adaptive LASSO and SCAD penalties},
	author={Fan, Jianqing and Feng, Yang and Wu, Yichao},
	journal={Annals of Applied Statistics},
	volume={3},
	number={2},
	pages={521--541},
	year={2009},
	publisher={Institute of Mathematical Statistics}
}

@article{li2019graphical,
	title={The graphical horseshoe estimator for inverse covariance matrices},
	author={Li, Yunfan and Craig, Bruce A and Bhadra, Anindya},
	journal={Journal of Computational and Graphical Statistics},
	volume={28},
	number={3},
	pages={747--757},
	year={2019},
	publisher={Taylor \& Francis}
}

@article{banerjee2015bayesian,
	title={Bayesian structure learning in graphical models},
	author={Banerjee, Sayantan and Ghosal, Subhashis},
	journal={Journal of Multivariate Analysis},
	volume={136},
	pages={147--162},
	year={2015},
	publisher={Elsevier}
}

@article{wang2012bayesian,
	title={Bayesian Graphical Lasso Models and Efficient Posterior Computation},
	author={Wang, Hao},
	journal={Bayesian Analysis},
	volume={7},
	number={4},
	pages={867--886},
	year={2012}
}

@article{atchade2019quasi,
	title={Quasi-Bayesian estimation of large Gaussian graphical models},
	author={Atchad{\'e}, Yves F},
	journal={Journal of Multivariate Analysis},
	volume={173},
	pages={656--671},
	year={2019},
	publisher={Elsevier}
}

@article{jankova2017honest,
	title={Honest confidence regions and optimality in high-dimensional precision matrix estimation},
	author={Jankov{\'a}, Jana and van de Geer, Sara},
	journal={Test},
	volume={26},
	pages={143--162},
	year={2017},
	publisher={Springer}
}

@article{cai2016estimating,
	title={Estimating structured high-dimensional covariance and precision matrices: Optimal rates and adaptive estimation},
	author={Cai, T Tony and Ren, Zhao and Zhou, Harrison H},
	journal={Electronic Journal of Statistics},
	volume={10},
	pages={1--59},
	year={2016}
}

@article{fan2016overview,
	title={An overview of the estimation of large covariance and precision matrices},
	author={Fan, Jianqing and Liao, Yuan and Liu, Han},
	journal={The Econometrics Journal},
	volume={19},
	number={1},
	pages={C1--C32},
	year={2016},
	publisher={Oxford University Press Oxford, UK}
}

@article{cai2011constrained,
	author = {Tonym Cai and Weidong Liu and Xi Luo},
	title = {A Constrained l-1 Minimization Approach to Sparse Precision Matrix Estimation},
	journal = {Journal of the American Statistical Association},
	volume = {106},
	number = {494},
	pages = {594--607},
	year = {2011},
	publisher = {Taylor \& Francis},
	doi = {10.1198/jasa.2011.tm10155}
}

@article{banerjee2021bayesian,
	title={Bayesian inference in high-dimensional models},
	author={Banerjee, Sayantan and Castillo, Isma{\"e}l and Ghosal, Subhashis},
	journal={arXiv preprint arXiv:2101.04491},
	year={2021}
}

@article{zhang2022contraction,
	title={Contraction of a quasi-Bayesian model with shrinkage priors in precision matrix estimation},
	author={Zhang, Ruoyang and Yao, Yisha and Ghosh, Malay},
	journal={Journal of Statistical Planning and Inference},
	volume={221},
	pages={154--171},
	year={2022},
	publisher={Elsevier}
}

@article{bellec2018slope,
	title={Slope meets lasso: improved oracle bounds and optimality},
	author={Bellec, Pierre C and Lecu{\'e}, Guillaume and Tsybakov, Alexandre B},
	journal={The Annals of Statistics},
	volume={46},
	number={6B},
	pages={3603--3642},
	year={2018},
	publisher={JSTOR}
}

@article{bhattacharya2016bayesian,
	title={Bayesian fractional posteriors},
	author={Bhattacharya, Anirban and Pati, Debdeep and Yang, YUN},
	journal={Annals of Statistics},
	volume={47},
	number={1},
	pages={39--66},
	year={2019},
	publisher={Institute of Mathematical Statistics}
}

@article{alquier2020concentration,
	title={Concentration of tempered posteriors and of their variational approximations},
	author={Alquier, Pierre and Ridgway, James},
	journal={The Annals of Statistics},
	volume={48},
	number={3},
	pages={1475--1497},
	year={2020}
}

@article{gao2020general,
	title={A general framework for Bayes structured linear models},
	author={Gao, Chao and van der Vaart, Aad W and Zhou, Harrison H},
	journal={Annals of Statistics},
	volume={48},
	number={5}, 
	pages={2848--2878},
	year={2020}
}

@article{sagar2024precision,
	title={Precision matrix estimation under the horseshoe-like prior--penalty dual},
	author={Sagar, Ksheera and Banerjee, Sayantan and Datta, Jyotishka and Bhadra, Anindya},
	journal={Electronic Journal of Statistics},
	volume={18},
	number={1},
	pages={1--46},
	year={2024},
	publisher={The Institute of Mathematical Statistics and the Bernoulli Society}
}

@article{meinshausen2006variable,
	title={Variable selection and high-dimensional graphs with the lasso},
	author={Meinshausen, Nicolai and B{\"u}hlmann, Peter},
	journal={Annals of Statistics},
	volume={34},
	pages={1436--1462},
	year={2006}
}

	\clearpage
\appendix

\begin{center}
{\huge\bfseries Supplementary Material for}\\[1em]
{\Large ``Censored Graphical Horseshoe: Bayesian sparse precision matrix estimation with censored and missing data"}
\\
The Tien Mai, Sayantan Banerjee
\end{center}

\section{Additional simulations results}
\label{app:addl_simu_results}

We now provide some trace and ACF plots as well as effective sample size of the Gibbs sampler to visualize the behaviour of our algorithm.

\begin{figure}[!ht]
	\centering
	\includegraphics[width=12cm]{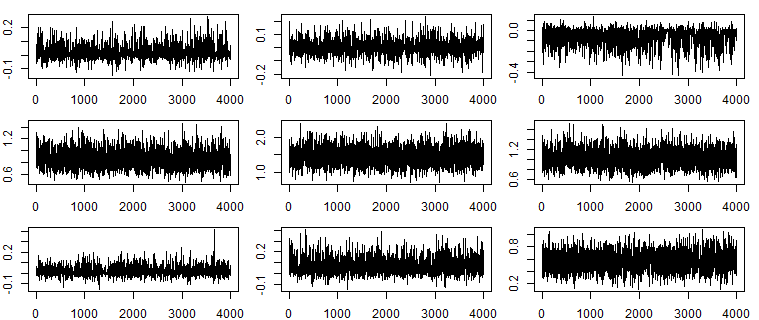}
	\caption{Trace plots from the Gibbs sampler for selected parameter entries.
		Top row: three randomly chosen entries with true value 0.
		Middle row: three randomly chosen entries with true value $1$.
		Bottom row: three randomly chosen entries with true value 0.3. The true precision matrix is in Setting I with $ p = 10, n = 100$. 10\% of data being missing.   }
	\label{fig_tracplot_I}
\end{figure}
\begin{figure}[!ht]
	\centering
	\includegraphics[width=14cm]{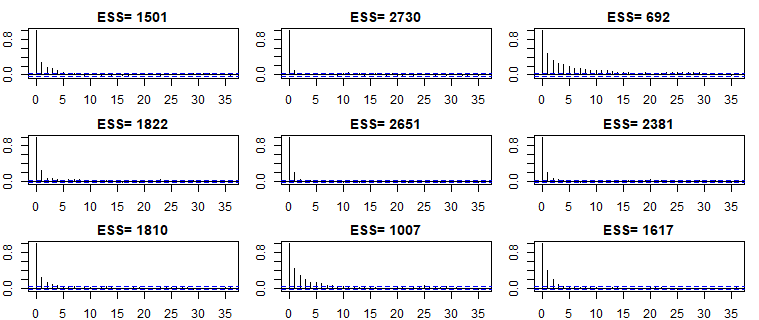}
	\caption{ACF plots from the Gibbs sampler for some random entries as in Figure \ref{fig_tracplot}. Top row (3 plots): 3 random entries with true value 0. Middle row (3 plots): 3 random entries with true value $1$. Bottom row (3 plots): 3 random entries with true value 0.3. The ESS (effective sample size) are also given.
    10\% of data being missing.}
	\label{fig_acf_plot_I}
\end{figure}

\begin{figure}[!ht]
	\centering
	\includegraphics[width=12cm]{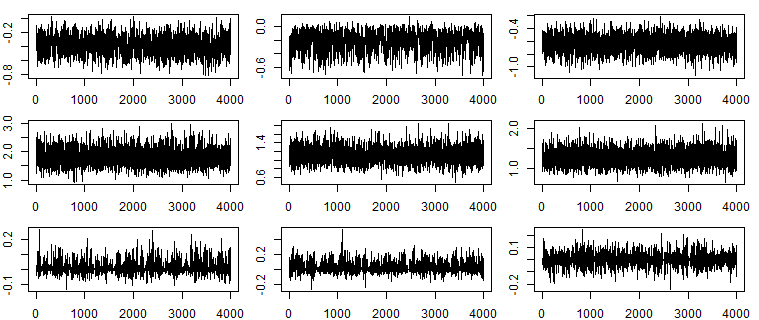}
	\caption{Trace plots from the Gibbs sampler for selected parameter entries.
		Top row: three randomly chosen entries with true value $-0.5$.
		Middle row: three randomly chosen entries with true value $1$.
		Bottom row: three randomly chosen entries with true value 0. The true precision matrix is in Setting II with $ p = 10, n = 100$. 10\% of data being missing.    }
	\label{fig_tracplot}
\end{figure}
\begin{figure}[!ht]
	\centering
	\includegraphics[width=14cm]{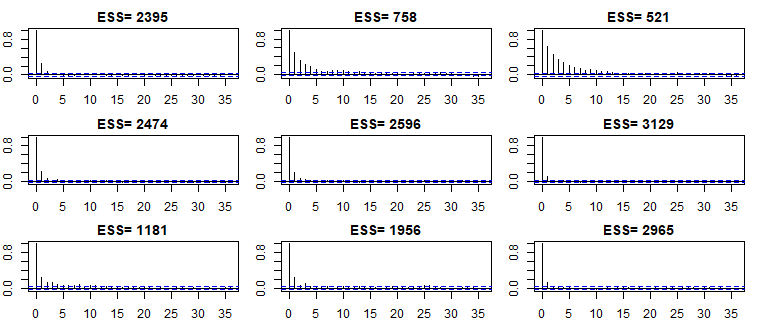}
	\caption{ACF plots from the Gibbs sampler for some random entries as in Figure \ref{fig_tracplot}. Top row: three randomly chosen entries with true value $-0.5$.
Middle row: three randomly chosen entries with true value $1$.Bottom row: three randomly chosen entries with true value 0. The true precision matrix is in Setting II with $ p = 10, n = 100$. The ESS (effective sample size) are also given. 10\% of data being missing.
    }
	\label{fig_acf_plot_II}
\end{figure}

\clearpage
\section{Proofs}
	\label{sc_proofs}
\subsection{Main proof}
In the proofs, $K$ represents a positive universal constant. Its value may vary from line to line, but we use it to simplify the presentation without specifying a new constant each time.

\sloppy

\begin{proof}[\textbf{Proof of Theorem \ref{thm_censoredGHS}}]
From Lemma \ref{lm_bounded_log_likelihood_censored}, we have that
$
 \mathcal{K}(P_{\Omega_0},P_{\Omega})   
 \leq
 K\,\|\Omega-\Omega_0\|_F^2.
$
Define 
$
\rho_n \propto \mathbf{1}_{\|\Omega - \Omega_0\|_2 < \delta } \Pi_{HS},$ 
where $\delta= [s\log (p/s)/n ]^{1/2} $. We have,
$$
		\int  \mathcal{K}(P_{\omega_0},P_{\omega}) \rho_n({\rm d}\omega)  
        \leq
        K \delta^2 = K \frac{s\log (p/s)}{n}.
		$$
From Lemma \ref{lm_bound_prior_horseshoe}, we have,
$$
\frac{1}{n}\mathcal{K}(\rho_n,\pi)  
    \leq
K \frac{s\log (p(p-1)/s)}{n}
    \lesssim
K \frac{s\log (p) }{n}
.
$$
Therefore, we can apply Theorem \ref{thm_expect_alquier} to have,
for any $\alpha\in(0,1)$,
		$$
		\mathbb{E} \left[ \int \mathcal{D}_{\alpha}(P_{\Omega},P_{\Omega_0}) \pi_{n,\alpha}({\rm d}\Omega ) \right]
		\leq \frac{1+\alpha}{1-\alpha}\varepsilon_n,
		$$
with $ \varepsilon_n = K n^{-1}s\log p  $, for some universal constant $K>0$ not depending on $n,p,s$. 
        
\end{proof}

\begin{proof}[\textbf{Proof of Theorem \ref{thm_missing_GHS}}]
From Lemma \ref{lm_bound_KL_for_missing}, we have that $
 \mathcal{K}(P_{\Omega_0},P_{\Omega})   
 \leq
K\,\|\Omega-\Omega_0\|_F^2.$ The rest of the proof is identical as above.       
\end{proof}

\subsection{Auxiliary results}

\begin{lemma}[Lemma 3 in \cite{mai2024concentration}]
		\label{lm_bound_prior_horseshoe}
Suppose $ \omega_0 \in \mathbb{R}^d $ such that $ \|\omega_0\|_0 = s^* $ and  that $ s^* < n < d $ and $ \|\omega_0\|_\infty \leq C_1  $. Suppose $\omega \sim \pi_{HS} $. 
Define $\delta_n =\{s^*\log (d/s^*)/n\}^{1/2} $. Then we have, for some constant $K>0 $, that
		$$
\pi_{HS} ( 
		\|\omega -\omega_0\|_2<\delta_n  )
		\geq 
		e^{-Ks^*\log (d/s^*)}.
		$$
and that $  \mathcal{K}(\rho_n,\pi_{HS})		\leq
\log \dfrac{1}{	\pi_{HS} (\|\omega - \omega_0 \|_2<\delta_n )}
		\leq
		K s^*\log (d/s^*)  $.
\end{lemma}	

We will make use the following general result, see also \cite{mai2025adaptive,mai2025high,mai2025handling,mai2025tobit,mai2025sparse}.

\begin{theorem}[Theorem 2.6 in \cite{alquier2020concentration}]
		\label{thm_expect_alquier}
Assume that there exists a sequence $\varepsilon_n > 0$ and a distribution $\rho_n$ such that
		$$
\int  \mathcal{K}(P_{\omega_0},P_{\omega}) \rho_n({\rm d}\omega)  \leq\varepsilon_n
		\text{; and } \,
\mathcal{K}(\rho_n,\pi)  \leq n\varepsilon_n.
		$$
		Then, for any $\alpha\in(0,1)$,
		$$
\mathbb{E} \left[ \int \mathcal{D}_{\alpha}(P_{\omega},P_{\omega_0}) \pi_{n,\alpha}({\rm d}\omega ) \right]
\leq 
\frac{1+\alpha}{1-\alpha}\varepsilon_n.
		$$
	\end{theorem}

\subsubsection{Lemma for censored data}

From \eqref{eq_likelihood_censored_data}, in the case of censored data,
the observed-data log-likelihood of $\widetilde y$ is as
$$
\ell(\Omega;\widetilde y)
=
\frac12\log\det(\Omega_{\mathcal O \mathcal O}) 
-
\frac12 \, \widetilde y_{\mathcal O}^\top \, \Omega_{\mathcal O \mathcal O} \, \widetilde y_{\mathcal O}
+ 
\log\Phi_{|\mathcal C|} \big(c_{\mathcal C};\,\mu_{\mathcal C\mid \mathcal O}(\widetilde y_{\mathcal O}),\Sigma_{\mathcal C\mid \mathcal O}\big),
$$
where $\mu_{\mathcal C\mid \mathcal O} (\widetilde y_{\mathcal O})
=
\Sigma_{\mathcal C \mathcal O} \Sigma_{\mathcal O \mathcal O}^{-1}\widetilde y_{\mathcal O } $ and $\Sigma_{\mathcal C\mid \mathcal O}
=
\Sigma_{\mathcal C \mathcal C}-\Sigma_{\mathcal C \mathcal O}\Sigma_{\mathcal O \mathcal O}^{-1}\Sigma_{\mathcal O \mathcal C}$.

\begin{lemma}
\label{lm_bounded_log_likelihood_censored}
Under Assumptions~\ref{assum_true_precision_Spectrum}
and \ref{assume_CDF_of_censored}, for the censored Gaussian model and log-likelihood $\ell(\Omega;\widetilde Y)$, 
there exists a constant $K>0 $ depending only on $m,M,$ and $c$, such that, for all $\Omega\in\mathcal S,$
$$
\mathbb{E}_{\Omega_0}\big[\ell(\Omega_0;\widetilde Y)-\ell(\Omega;\widetilde Y)\big] 
\le 
K\,\|\Omega-\Omega_0\|_F^2
.
$$
\end{lemma}

\begin{proof}[\textbf{Proof of Lemma \ref{lm_bounded_log_likelihood_censored}}]

\textit{Step 1: Taylor Expansion of the Expected Log-Likelihood.}
Let $L(\Omega) := \mathbb{E}_{\Omega_0}[\ell(\Omega; \widetilde Y)]$ denote the expected log-likelihood under the true precision matrix $\Omega_0$. Since the log-likelihood function $\ell(\Omega; \widetilde Y)$ is twice continuously differentiable with respect to $\Omega$ on the spectral band $\mathcal{S}$, we apply a second-order Taylor expansion of $L(\Omega)$ around $\Omega_0$. 

Using the Lagrange form of the remainder, there exists some $\Omega^* = \Omega_0 + t(\Omega - \Omega_0)$ with $t \in (0,1)$ such that:
$$
L(\Omega) - L(\Omega_0) = \langle \Omega - \Omega_0, \nabla L(\Omega_0) \rangle + \frac{1}{2} \langle \Omega - \Omega_0, \nabla^2 L(\Omega^*)[\Omega - \Omega_0] \rangle,
$$
where $\langle A, B \rangle = \text{tr}(A^\top B)$ denotes the Frobenius inner product.

\textit{Step 2: The First-Order Term Vanishes.}
The gradient of the expected log-likelihood, $\nabla L(\Omega)$, is given by $\mathbb{E}_{\Omega_0}[\nabla \ell(\Omega; \widetilde Y)]$. A fundamental property of maximum likelihood estimation is that the expected score at the true parameter is zero. Thus:
$$
\nabla L(\Omega_0) = \mathbb{E}_{\Omega_0}[\nabla \ell(\Omega_0; \widetilde Y)] = 0.
$$
Substituting this into the expansion yields:
$$
L(\Omega) - L(\Omega_0) = \frac{1}{2} \langle \Omega - \Omega_0, \nabla^2 L(\Omega^*)[\Omega - \Omega_0] \rangle.
$$
Rearranging to match the quantity of interest:
$$
\mathbb{E}_{\Omega_0}\big[\ell(\Omega_0; \widetilde Y) - \ell(\Omega; \widetilde Y)\big] = L(\Omega_0) - L(\Omega) = -\frac{1}{2} \langle \Omega - \Omega_0, \nabla^2 L(\Omega^*)[\Omega - \Omega_0] \rangle.
$$

\textit{Step 3: Uniform Bound on the Hessian.}
Since $\mathcal{S}$ is convex, the intermediate matrix $\Omega^*$ lies within $\mathcal{S}$. By Assumption \ref{assume_CDF_of_censored}, the Hessian of the expected log-likelihood is uniformly bounded on $\mathcal{S}$. Specifically, for any matrix $\Delta$, the quadratic form satisfies:
$$
\big| \langle \Delta, \nabla^2 L(\Omega^*)[\Delta] \rangle \big| \le K_1 \|\Delta\|_F^2,
$$
for some constant $K_1 > 0$ depending on $m, M,$ and $c$.

Since the expected log-likelihood is concave, the term $\langle \Omega - \Omega_0, \nabla^2 L(\Omega^*)[\Omega - \Omega_0] \rangle$ is non-positive. Therefore:
$$
-\frac{1}{2} \langle \Omega - \Omega_0, \nabla^2 L(\Omega^*)[\Omega - \Omega_0] \rangle 
\leq
\frac{1}{2} \big| \langle \Omega - \Omega_0, \nabla^2 L(\Omega^*)[\Omega - \Omega_0] \rangle \big|
\le
\frac{1}{2} K_1 \|\Omega - \Omega_0\|_F^2.
$$
Setting $K = \frac{1}{2} K_1$, we obtain the desired result:
$$
\mathbb{E}_{\Omega_0}\big[\ell(\Omega_0; \widetilde Y) - \ell(\Omega; \widetilde Y)\big] \le K \|\Omega - \Omega_0\|_F^2.
$$
The constant $K > 0$ depends on $m, M,$ and the censoring thresholds $c$. The proof is completed.

\end{proof}

\subsubsection{Lemma for missing data}
We assume that the (possibly random) pattern $(\mathcal O_i)_{i=1}^n$ is arbitrary but independent of the unobserved values (MAR). From \eqref{eq_likelihood_missing}, for the case of missing data, we have the log-likelihood as
$$
\ell_i(\Omega;\widetilde Y_i) 
\;=\; -\frac12\log\det(\Sigma_{\mathcal O_i})
-\frac12 \widetilde Y_i^\top \Sigma_{\mathcal O_i}^{-1}\widetilde Y_i.
$$

\begin{lemma}
\label{lm_bound_KL_for_missing}
Under Assumption \ref{assum_true_precision_Spectrum}, there exists a finite constant $ K >0$, depending only on $m$ and $M$, such that,
$$
E_{\Omega_0}\big[\,\ell(\Omega_0;\widetilde Y)-\ell(\Omega;\widetilde Y)\,\big]
\le
K\,\|\Omega-\Omega_0\|_F^2.
\;
$$
\end{lemma}

\begin{proof}[\textbf{Proof of Lemma \ref{lm_bound_KL_for_missing}}]    

We prove the statement step by step and track constants.

\textit{Step 1.} Fix a row $i$ and its observed index set $O:=\mathcal O_i$. Let $A:=\Sigma_{0,O}$ and $B:=\Sigma_{O}$ denote the corresponding principal submatrices of $\Sigma_0$ and $\Sigma$. Conditional on the pattern, $\widetilde Y_i\sim\mathcal N_d(0,A)$ under the true model ($d:=|O|$). The expected contribution of row $i$  equals the KL divergence between these two centered Gaussians:
$$
\delta_i \;:=\; \mathbb E_{\Omega_0}\big[\ell_i(\Omega_0;\widetilde Y_i)-\ell_i(\Omega;\widetilde Y_i)\big]
= \mathcal{K} \big(\mathcal N(0,A)\,\big\|\,\mathcal N(0,B)\big).
$$
The standard closed form gives
$$
\delta_i \;=\; \frac12\Big\{\operatorname{tr}(B^{-1}A)-d + \log\frac{\det B}{\det A}\Big\}.
$$

\textit{Step 2.} 
Set $E:=A^{-1/2} B A^{-1/2}$ (so $E\succ0$). Then
$$
\delta_i 
=
\frac12\Big\{\operatorname{tr}(E^{-1})-d + \log\det E\Big\} 
=:
\frac12\phi(E).
$$
Note $\phi(I)=0$ and $\nabla\phi(I)=0$ (first-order term cancels). On the spectral bounds from Assumption \ref{lm_bound_KL_for_missing}, we have
$$
\lambda_{\min}(A)\ge \lambda_{\min}(\Sigma_0)\ge 1/M,\qquad
\lambda_{\max}(A)\le \lambda_{\max}(\Sigma_0)\le 1/m,
$$
and similarly for $B$. Hence the eigenvalues of $E$ lie in the compact interval $[m/M,\,M/m]$. Let $\mathcal S$ denote the set of such $E$ (with dimension at most $p$). The function $\phi$ is twice continuously differentiable on $\mathcal S$, hence its Hessian is bounded there; write
$$
H_{\max}:=\sup_{E\in\mathcal S}\|{\rm Hess}\,\phi(E)\|_{\mathrm{op}}<\infty.
$$
By the second-order Taylor formula with integral remainder (or the mean-value form of the remainder) applied at $I$,
$$
\phi(E) \;=\; \tfrac12 \langle E-I,\; {\rm Hess}\,\phi(\xi)[E-I]\rangle
\le \tfrac12 H_{\max}\|E-I\|_F^2
$$
for some $\xi$ on the segment between $I$ and $E$. Consequently
$$
\delta_i \;=\; \tfrac12\phi(E) \;\le\; \tfrac{H_{\max}}{4}\,\|E-I\|_F^2.
$$

\textit{Step 3.} Observe that $E-I = A^{-1/2}(B-A)A^{-1/2}$. Hence
$$
\|E-I\|_F \le \|A^{-1/2}\|_2^2\,\|B-A\|_F = \frac{1}{\lambda_{\min}(A)}\|B-A\|_F.
$$
Because $\lambda_{\min}(A)\ge 1/M$ we get $\|E-I\|_F \le M\,\|B-A\|_F$. Therefore
$$
\delta_i \le \frac{H_{\max}}{4}\,M^2\,\|B-A\|_F^2.
$$

\textit{Step 4.} We use the exact identity
$$
\Sigma-\Sigma_0 \;=\; \Omega^{-1} - \Omega_0^{-1} \;=\; \Omega^{-1}(\Omega_0-\Omega)\Omega_0^{-1}.
$$
Therefore with the operator-norm bounds $\|\Omega^{-1}\|_2,\|\Omega_0^{-1}\|_2\le 1/m$,
$$
\|\Delta\|_F \le \|\Omega^{-1}\|_2\,\|\Omega_0^{-1}\|_2\,\|\Omega-\Omega_0\|_F
\le \frac{1}{m^2}\,\|\Omega-\Omega_0\|_F.
$$
Combining with the previous step yields
$$
\Delta(\Omega) 
\le 
\frac{H_{\max}}{4} \,\frac{M^2}{m^4}\,\|\Omega-\Omega_0\|_F^2.
$$
This proves the claimed upper bound. 
\end{proof}

\end{document}